\documentclass{article}
\usepackage{emulateapj}
\usepackage{apjfonts}

\begin{document}

\submitted{To appear in The Astrophysical Journal Letters}

\title{Evidence for a Significant Intermediate-Age Population
in the M31 Halo from Main Sequence Photometry$^1$}

\author{Thomas M. Brown, Henry C. Ferguson, Ed Smith}

\affil{Space Telescope Science Institute, 3700 San Martin Drive,
Baltimore, MD 21218;  tbrown@stsci.edu, ferguson@stsci.edu,
edsmith@stsci.edu} 

\medskip

\author{Randy A. Kimble, Allen V. Sweigart}

\affil{Code 681, NASA Goddard Space Flight Center, Greenbelt, MD 20771; 
randy.a.kimble@nasa.gov, allen.v.sweigart@nasa.gov} 

\author{Alvio Renzini}

\affil{European Southern Observatory, Karl-Schwarzschild-Strasse 2, 
Garching bei M$\ddot{\rm u}$nchen, Germany; arenzini@eso.org} 

\author{R. Michael Rich}

\affil{Division of Astronomy, Dpt.\ of Physics \& Astronomy, UCLA, Los
Angeles, CA 90095; rmr@astro.ucla.edu}

\author{Don A. VandenBerg}

\affil{Department of Physics and Astronomy, University of Victoria, P.O. Box
3055, Victoria, BC, V8W 3P6, Canada; davb@uvvm.uvic.ca}

\begin{abstract}

We present a color-magnitude diagram (CMD) for a minor-axis field in
the halo of the Andromeda galaxy (M31), 51 arcmin (11 kpc) from the
nucleus.  These observations, taken with the Advanced Camera for
Surveys (ACS) on the Hubble Space Telescope, are the deepest optical
images yet obtained, attaining 50\% completeness at $m_V =
30.7$~mag. The CMD, constructed from $\sim 3 \times 10^5$ stars, 
reaches more than 1.5~mag fainter than the old main-sequence
turnoff.  Our analysis is based on direct comparisons to ACS
observations of four globular clusters through the same
filters, as well as $\chi^2$ fitting to a finely-spaced grid of
calibrated stellar-population models. We find that the
M31 halo contains a major ($\sim 30$\% by mass) intermediate-age (6--8
Gyr) metal-rich ($ \rm [Fe/H] > -0.5$) population, as well as a
significant globular-cluster age (11--13.5 Gyr) metal-poor
population.  These findings
support the idea that galaxy mergers played an important role
in the formation of the M31 halo.

\end{abstract}


\section{Introduction}

A primary quest of observational astronomy is to establish the
formation history of galaxies.  Studies of the stellar populations
at the main-sequence turnoff
(MSTO) provide the most precise measurements, but
these generally have been limited to the Galaxy and its satellites.
The Galaxy has driven much of our thinking regarding the formation
of giant spirals, with its predominately old metal-poor halo (e.g.,
VandenBerg 2000; Ryan \& Norris 1991) and younger disk (e.g.,
Sourbiran, Bienaym$\rm \acute{e}$, \& Siebert 2003; Fontaine,
Brassard, \& Bergeron 2001).  However, an important contrast is
offered by our nearest giant neighbor, Andromeda (M31; NGC224).
Although little is known about the formation history of M31, its halo
has a strikingly higher metallicity (Mould \& Kristian 1986; Holland,
Fahlman, \& Richer 1996; Durrell, Harris, \& Pritchet 2001), despite
its similar Hubble type to the Galaxy.

Hierarchical models suggest that spheroids (bulges
and halos) form in a repetitive process during the mergers of galaxies
and protogalaxies, while disks form by slow accretion
of gas between merging events (e.g., White \& Frenk
1991).  The discovery of the Sgr dwarf galaxy (Ibata et al.\ 1994) 
sparked renewed interest in halo
formation through accretion of dwarf galaxies.  Ambitious programs are
now mapping the spatial distribution, kinematics, and chemical
abundance in the halos of the Galaxy (e.g., Morrison
\linebreak

{\small \noindent $^1$Based on observations made with the NASA/ESA
Hubble Space Telescope, obtained at the Space Telescope Science Institute, 
which is operated by AURA, Inc., under NASA contract NAS 5-26555. These
observations are associated with proposal 9453.}

\noindent
 et al.\ 2000;
Majewski et al.\ 2000) and M31 (e.g., Ferguson et al.\ 2002).  In the
meantime, the overproduction of dwarf galaxies in hierarchical models
has led to suggestions that most of the dwarfs formed in the early
Universe have dissolved into the halo (e.g., Bullock et al.\ 2000).
Whether or not such accretion is the dominant source of halo stars, it
is likely that dwarf galaxies do contribute, with their stars
sometimes remaining in coherent orbital streams for many Gyr.  Indeed,
such substructure has been found in the M31 halo (Ferguson et al.\ 2002).

With the installation of the Advanced Camera for Surveys (ACS; Ford et
al. 1998) on the Hubble Space Telescope (HST), it is now possible to
resolve the old main sequence population in the M31 halo and analyze
the age distribution via the same techniques that have been applied to
Galactic globular clusters (GCs) and satellite dwarf galaxies.  M31
offers a unique testing ground for understanding giant galaxy
formation, due to its proximity (770 kpc; Freedman \& Madore
1990), small foreground reddening ($E_{B-V}=0.08$~mag; Schlegel,
Finkbeiner, \& Davis 1998), and low inclination ($12.5^{\rm o}$; de
Vaucouleurs 1958).  To investigate the star formation history of the
M31 halo, we have obtained deep ACS observations of a minor axis field
$\approx 51$ arcmin (11 kpc) from the nucleus.  In this initial report,
we briefly describe the observations and report that the
halo contains a significant population of intermediate age (6--8 Gyr)
metal-rich ($\rm [Fe/H] > -0.5$) stars.

\section{Observations and Data Reduction}

Using the ACS Wide Field Camera, we obtained deep optical images of a
field along the SE minor axis of the M31 halo, at
$\alpha_{2000} = 00^h46^m07^s$, $\delta_{2000} = 40^{\rm
o}42^{\prime}34^{\prime\prime}$.  The field is not associated with the
tidal streams and substructure found by Ferguson et al.\ (2002), and
lies just outside the ``flattened inner halo'' in their maps.  The
field was previously imaged by Holland et al.\ (1996),
using the Wide Field Planetary Camera 2 (WFPC2).  Given the highly
inclined disk, the contribution of disk stars at this position should
be $\lesssim 3$\% (Walterbos \& Kennicutt 1988; Holland et al.\
1996 and references therein). We chose this field to optimize the
crowding (trading off population statistics versus photometric
accuracy) and to place an interesting M31 GC (GC312; Sargent et al.\
1977) near the edge of our images; GC312 will be analyzed in a
future paper.  

From 2 Dec 2002 to 11 Jan 2003, we obtained 39.1 hours of images in
the F606W filter (broad $V$) and 45.4 hours in the F814W filter ($I$),
with every exposure dithered to allow for hot pixel removal, optimal point
spread function sampling, smoothing of spatial variation in detector
response, and filling in the gap between the two halves of the $4096
\times 4096$ pix detector.  We also obtained parallel WFPC2 observations
of a field further out in the M31 halo, using the same
filters; the WFPC2 data are much less deep than the ACS
data, but they reach the MSTO, and will be
analyzed in a forthcoming paper.

We co-added the M31 images using the IRAF DRIZZLE package, with masks
for the cosmic rays and hot pixels, resulting in geometrically-correct
images with a plate scale of 0.03$^{\prime\prime}$ pixel$^{-1}$ and an
area of approximately $210^{\prime\prime} \times 220^{\prime\prime}$.
We then performed both aperture and PSF-fitting photometry using the
DAOPHOT-II package (Stetson 1987), assuming a variable PSF constructed
from the most isolated stars.  The aperture photometry on
isolated stars was corrected to true apparent magnitudes using TinyTim
models of the HST PSF (Krist 1995) and observations of the standard star
EGGR 102 (a $V=12.8$~mag DA white dwarf) 
in the same filters, with agreement at the 1\% level.  The
PSF-fitting photometry was then compared to the corrected aperture
photometry, in order to derive the offset between the PSF-fitting
photometry and true apparent magnitudes. Our photometry is in the STMAG 
system: $m= -2.5 \times $~log$_{10}
f_\lambda -21.1$.  For readers more familiar with the 
Johnson $V$ and Cousins $I$ bandpasses, a 5,000~K stellar spectrum
has $V - m_{F606W} = -0.05$~mag and $I - m_{F814W} = -1.28$~mag.

Of the $\sim$300,000 stars detected, we discarded those within the GC312 tidal
radius (10$^{\prime\prime}$; Holland et al.\ 1997), within
14.5$^{\prime\prime}$ of a bright foreground star, within
12.6$^{\prime\prime}$ of this star's window reflection, and near the
image edges, leaving $\approx$223,000 stars in the final catalog.  
Using the SExtractor code (Bertin \& Arnouts 1996), we
estimate $\lesssim 5$\% of the stars are contaminated by extended
sources.  We show the CMD in Figure 1$a$.
Extensive artificial star tests determine the photometric scatter and
completeness as a function of color and luminosity.  The CMD shows no
obvious differences when comparing the population in a
10--100$\arcsec$ annulus around GC312 to that beyond 100$\arcsec$; the
cluster does not appear to be associated with an extended underlying
system.  By integrating our catalog, we estimate that the surface
brightness in our field is $\mu_V \approx 26.3$~mag arcsec$^{-2}$.

Our program includes ACS observations of four Galactic GCs spanning a
wide metallicity range (Table 1), using the same filters, in order to
obtain fiducials in these bands and to verify the transformation of
theoretical isochrones to the observational plane.  M92, NGC6752, and
47~Tuc are the most useful calibrators in our program, because their
parameters are known very well;
NGC6528 is also useful because of its high metallicity, but its
parameters are less secure, and it suffers from high, spatially variable
reddening (Heitsch \& Richtler 1999).  The GC images were
not dithered, so these data were drizzled without plate scale changes,
in order to remove cosmic rays and to correct for geometric
distortion.  These images are significantly less crowded than those
of M31, and the PSF is undersampled, so we performed aperture
photometry but no PSF fitting, and corrected the aperture photometry
to true apparent magnitudes. \\

\noindent
\parbox{3.0in}{
{\sc Table 1:} M31 and globular cluster parameters

\begin{tabular}{cccc}
\tableline
Name    & $(m-M)_V$ (mag) & $E_{B-V}$ (mag) & [Fe/H] \\
\tableline
M31     & 24.68$^a$ & 0.08$^b$  & $-0.6^c$ \\
\tableline
M92     & 14.60$^d$ & 0.023$^b$ & $-2.14^d$ \\
NGC6752 & 13.17$^e$ & 0.055$^b$ & $-1.54^f$ \\
47~Tuc   & 13.27$^g$ & 0.032$^b$ & $-0.83^f$ \\
NGC6528 & 16.15$^h$ & 0.55$^h$  & $-0.2^h$ \\
\tableline
\end{tabular}
$^a$Freedman \& Madore (1990).\\
$^b$Schlegel et al.\ (1998).\\
$^c$Mould \& Kristian (1986).\\
$^d$VandenBerg \& Clem (2003).\\
$^e$Renzini et al.\ (1996).\\
$^f$VandenBerg (2000).\\
$^g$Zoccali et al.\ (2001).\\
$^h$Momany et al.\ (2003).
}

\section{Analysis}

Our CMD reveals, for the first time, the main sequence
population in the M31 halo. The horizontal branch (HB) extends from a
well-populated red clump to a minority blue population ($\sim 10$\%).
The red giant branch (RGB) shows the broad color
distribution extending to high metallicity, long-known to be
characteristic of the M31 halo (Mould \& Kristian 1986).  The
luminosity difference between the red HB and the RGB ``bump'' reaches
0.5~mag -- another indication of near-solar metallicities.  There is
also a prominent blue plume of stars significantly brighter than the
MSTO; this minority population ($\sim 2$\% the size of the population
at the MSTO $\pm$1~mag) may include binaries, blue stragglers, or a
residual young stellar population.  Although 
the blue HB stars are characteristic of very old, metal-poor
GCs (such as M92), the luminosity difference between the MSTO and HB is
smaller than expected for a purely old stellar population.  This is
shown in Figure 1$b$, which includes the ridge lines and HB loci for
the four GCs we observed with ACS.  Also, the M31 subgiant
branch is nearly horizontal, indicating a high metallicity,
while its ridge is appreciably brighter than that of 47~Tuc and NGC6528,
implying the presence of a significantly younger population in the
M31 halo.

Our full analysis of the star formation history will be presented in a
future paper. Here, we will supplement the comparisons shown in Figure
1$b$ with examples from our modeling to date.  Our modeling is based
upon the isochrones of VandenBerg et al.\ (2003), which show good
agreement with cluster CMDs spanning a wide range of metallicity and
age.  To transform these isochrones from the ground-based to HST-based
bandpasses, we used the spectra of Lejeune, Cuisinier, \& Buser (1997)
to calculate $V - m_{F606W}$ and $I - m_{F814W}$, then applied those
differences to the ground-based magnitudes of the isochrones, with a
small ($\lesssim 0.05$~mag) empirical color correction to force
agreement with our GC CMDs. After this correction, 
the isochrones match the ridge lines within $\lesssim 0.02$~mag
over the region of the CMD we are fitting. Our observed CMDs
of these clusters are reproduced by a 12.5~Gyr isochrone for 
47~Tuc and 14~Gyr isochrones for NGC6752 and M92.  The 
isochrones do not include core He diffusion, which would reduce their ages by
10--12\%, 
\linebreak

\parbox{6.5in}{\small {\sc Fig.~1--} {\it (a)} The CMD of M31's halo,
shown as a Hess diagram with a logarithmic stretch and completeness
limits ({\it green and blue lines}) labeled.  {\it (b)} The M31 CMD
with the ridge lines ({\it colored curves}) and HB stars ({\it colored
points}) of 4 GCs overplotted.  Except for NGC6528, the GCs have only
been shifted by the differences in distance and reddening (Table 1).
The NGC6528 ridge line and HB were shifted 0.16~mag brighter to align
its HB to that of M31 (NGC6528's distance and reddening are very uncertain).
Note that much of the M31 subgiant branch is brighter than those of
the clusters, indicating that M31's halo is predominantly younger than
Galactic GCs. {\it (c)} The region of the M31 CMD used for fitting the
population age and metallicity distributions.  The histogram ({\it
inset, red}) shows the number of stars along a cut ({\it red line})
through the subgiant branch.  {\it (d)} The best fit achieved using
old (11--13.5~Gyr) isochrones spanning a range in metallicity ($-2.31
< \rm [Fe/H] < 0$).  The inset compares the histogram ({\it black})
across the subgiant branch ({\it black line}) to that from the
previous panel ({\it red}).  Although the RGB width requires a spread
in metallicity, old stars by themselves cannot reproduce the M31 CMD
below the RGB.  {\it (e)} The CMD for the best-fit population, with a
significant spread in both age and metallicity.  The two dominant
components to the fit are a majority population of intermediate-age
(6--11~Gyr) metal-rich ([Fe/H]~$> -0.5$) stars, and a smaller
population of old (10--13.5~Gyr) metal-poor ([Fe/H]~$< -0.5$)
stars. The inset compares a cut across the subgiant branch in the
model ({\it black}) to that in the data ({\it red}), showing good
agreement.  {\it (f)} The same best-fit model, color-coded to highlight the
intermediate-age metal-rich ({\it red}) and old metal-poor ({\it
blue}) components.}

\epsfxsize=6.5in \epsfbox{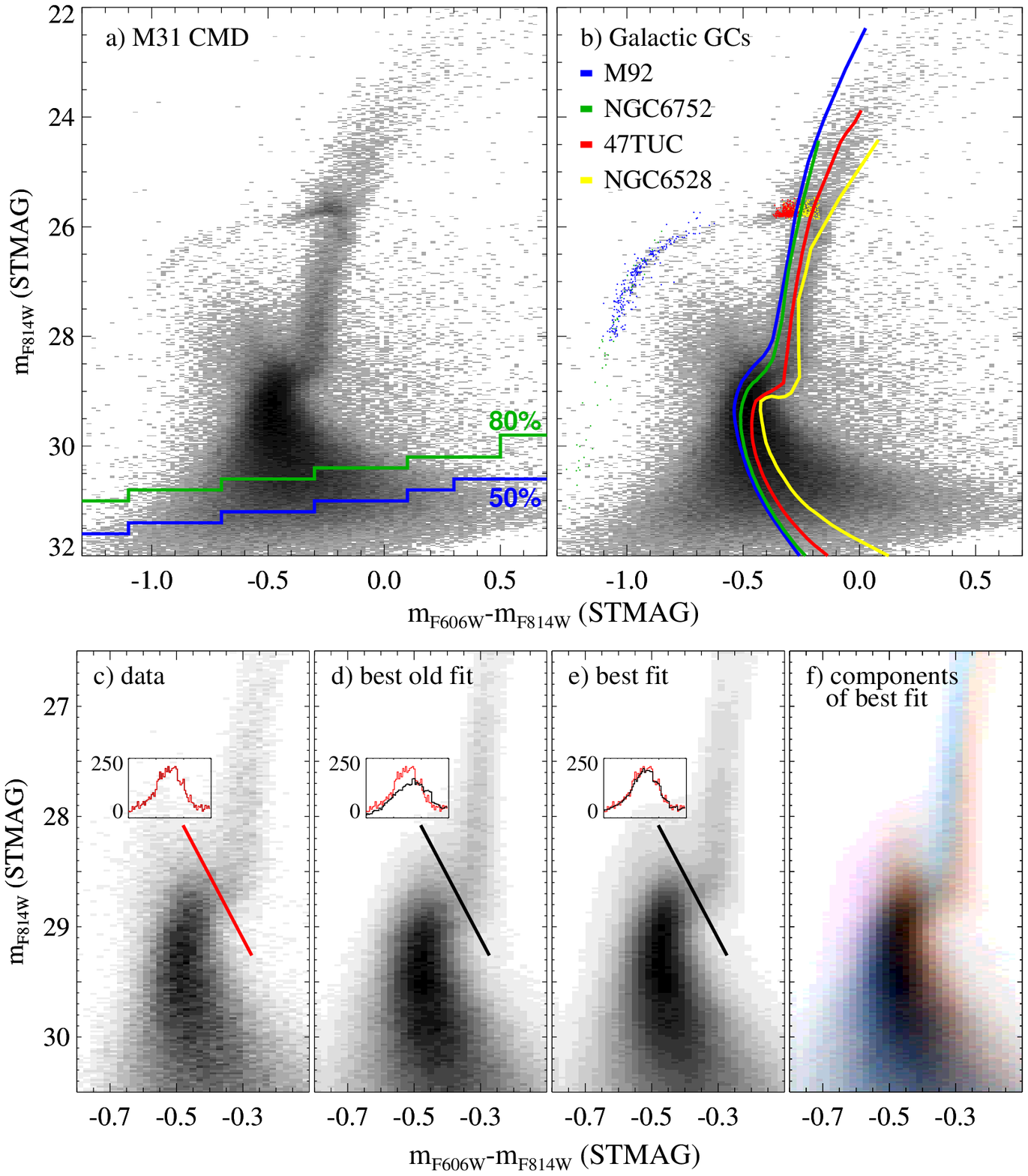} 

\newpage

\noindent
thus avoiding discrepancies with the age of the Universe
(VandenBerg et al.\ 2002).

Using isochrones with a range of ages and metallicities, we fit the
region of the M31 CMD shown in Figure 1$c$ using the StarFish code of
Harris \& Zaritsky (2001).  Restricting the fit to this region of the
CMD focuses on the most sensitive age and metallicity indicators while
avoiding regions of the CMD that are poorly constrained by
the models (e.g., the HB morphology).  The StarFish code
fits the observed CMD through a linear combination of input
isochrones, using $\chi^2$ minimization, where the isochrones are
scattered according to the results of the artificial star tests.  We
first followed the standard method for determining the metallicity
distribution from the RGB with a fit (not shown) to the RGB using a
set of old (13.5~Gyr) isochrones spanning a wide metallicity range
($-2.31 < \rm [Fe/H] < 0$).  The resulting metallicity distribution
was similar to that found by Holland et al.\ (1996) in our same field,
and Durrell et al.\ (2001) in a field 20 kpc from the nucleus.
However, it was very obvious that the isochrones did not match the
subgiant branch or the MSTO.  Next, as shown in Figure 1$d$, we tried
fitting the entire region of Figure 1$c$ with a wider age range
(11.5--13.5 Gyr), but found no acceptable combination.  The insets of
Figure 1$c$, $d$, and $e$ are histograms showing the number of stars
(data: {\it red}; model: {\it black}) along a cut through the subgiant
branch ({\it thick line}). In Figure 1$d$, the residual subgiant stars that
are not reproduced by this old model ({\it inset}) suggest that at
least 20\% of the stars in the halo must be younger than 11~Gyr.
We conclude that a purely old stellar population cannot explain the
CMD of the M31 halo.

Next, we expanded the age range to 6--13.5~Gyr and repeated the fit
(see Figure 1$e$).  The width of the RGB is now matched without a
mismatch at the subgiant branch ({\it inset}).  In Figure 1$f$, we
highlight the two dominant populations in the best-fit model: 56\% of
the population ({\it red}) is metal-rich ([Fe/H]$> -0.5$) and of
intermediate age (6--11 Gyr), while 30\% of the population ({\it
blue}) is metal-poor ([Fe/H]$< -0.5$) and old (11--13.5 Gyr).  About
half of this metal-rich population (i.e., 28\% of the total
population) is 6--8~Gyr old.  We stress that these models only
illustrate, in a broad sense, the dominant populations present in the
M31 halo.  Other combinations of young metal-rich and old metal-poor
stars are possible.  For example, we produced a similar fit by
combining two very distinct isochrone groups: one at 6--8 Gyr with
[Fe/H]$>-1$, and one at 11.5--13.5 Gyr with [Fe/H]$<-1$.  In a future
paper, we will present more detailed constraints on the star formation
history.

\section{Summary and Discussion}

The CMD of the M31 halo is evidently inconsistent with a population
composed solely of old (GC-age) stars; instead, it is dominated by a
population of metal-rich intermediate-age stars.  Although the high
metallicity in the M31 halo is well-documented, the large age spread
required to simultaneously reproduce the RGB, subgiant, and main sequence
distributions came to us as a surprise. Earlier studies of the RGB
were insensitive to this age spread.  For example, Durrell et al.\
(2001) were able to explain the metallicity distribution 20 kpc from
the nucleus, with a simple chemical evolution model forming most of
the stars at very early times (see also C$\rm \hat{o}$t$\rm \acute{e}$
et al.\ 2000). Although our field is relatively small in sky coverage,
it appears representative; the metallicity in our field (Holland et
al.\ 1996) agrees well with that much further out (Durrell et al.\
2001), and there are no indications of substructure or tidal streams
in the region we surveyed (Ferguson et al.\ 2002).

It seems unlikely that star formation in the halo proceeded for
$\sim$6~Gyr from gas in situ; instead, the broad age dispersion in the
halo is likely due to contamination from the disruption of satellites
or of disk material into the halo during mergers.  Indeed, our current
analysis of the data cannot rule out the possibility that M31 and a
nearly-equal-mass companion galaxy experienced a violent merger when
the Universe was half its present age.  If the 6--8 Gyr population in
the halo represents the remnants of a disrupted satellite, the
relatively high metallicity suggests that it must have been fairly massive.
On the other hand, the stars could represent
disruption of the M31 disk, either by a major collision when M31 was
$\sim 6$ Gyr old, or by repeated encounters with smaller satellites.
The resulting halo would be a mix of the old metal-poor stars formed
earliest in M31's halo, disk stars that formed prior to the merger(s),
stars formed during the merger(s), and the remnant populations of the
disrupted satellite(s).
 
\acknowledgements

Support for proposal 9453 was provided by NASA through a grant from
STScI, which is operated by AURA,
Inc., under NASA contract NAS 5-26555.  We are grateful to J.\ Harris 
and P.\ Stetson for providing their codes and assistance.
We thank the members of the scheduling
and operations teams at STScI
(especially P.\ Royle, D.\ Taylor, and D.\ Soderblom) for
their efforts in executing a large program during a busy HST cycle.

\end{document}